\documentclass[aps, prl, superscriptaddress, nofootinbib, twocolumn]{revtex4}

\usepackage{amsmath}	% AMS Math Package
\usepackage{amsthm}	% Theorem Formatting
\usepackage{amssymb}	% Math symbols such as \mathbb
\usepackage{bbm} % font for double-lined letters _and_ numbers, e.g. matrix one; usage: \mathbbm{...}
\usepackage{datetime}
\usepackage{graphicx}
\usepackage{color}
\usepackage{verbatim}
\usepackage[colorlinks=true, citecolor=midblue, linkcolor=midblue, urlcolor=midblue]{hyperref}

\definecolor{grey}{rgb}{0.4,0.4,0.4}
\definecolor{dullmagenta}{rgb}{0.4,0,0.4}
\definecolor{darkblue}{rgb}{0,0,0.4}
\definecolor{midblue}{rgb}{0,0,0.5}
\definecolor{midred}{rgb}{0.5,0,0}
\definecolor{orange}{rgb}{1,0.5,0}
\definecolor{lightbrown}{rgb}{0.75,0.5,0.25}
\definecolor{tan}{cmyk}{0.14,0.42,0.56,0}
\definecolor{djunglegreen}{cmyk}{0.99,0,0.52,0}
\definecolor{lightgreen}{rgb}{0,1,0}
\definecolor{olivegreen}{cmyk}{0.64,0,0.95,0.40}
\definecolor{midgreen}{rgb}{0.0,0.675,0.0}
\definecolor{darkgreen}{rgb}{0,0.5,0}

\newcommand{\normdouble}[1]{\left\|{#1}\right\|}
\newcommand{\norm}[1]{\normdouble{#1}}

\newcommand{\vs}{\vspace}

\renewcommand{\.}{\hspace{0.5mm}}

\newcommand{\Arm}{\ensuremath{\mathrm{A}}}
\newcommand{\Brm}{\ensuremath{\mathrm{B}}}

\newcommand{\brm}{\ensuremath{\mathrm{b}}}
\newcommand{\crm}{\ensuremath{\mathrm{c}}}

\newcommand{\rrm}{\ensuremath{\mathrm{r}}}

\newcommand{\Pcal}{\ensuremath{\mathcal{P}}}

\newcommand{\Rcal}{\ensuremath{\mathcal{R}}}

\renewcommand{\d}{\ensuremath{\mathrm{d}}}
\newcommand{\ie}{i.e.}

\newcommand{\cf}{cf.} % CR: according to Oxford English, ``c.f.'' means carried forward

\let\baraccent=\= % rename builtin command \= to \baraccent
\renewcommand{\=}[1]{\stackrel{#1}{=}} % for putting numbers above =

\theoremstyle{definition}

\theoremstyle{remark}

\newcommand{\Msun}{M_{\odot}}

\smallskipamount
\settimeformat{ampmtime}

\usepackage{float}

\begin{document}

\title{Enhanced Detectability of Spinning Primordial Black Holes}

\author{Florian K{\"u}hnel}
\email{florian.kuhnel@fysik.su.se\\}
\affiliation{The Oskar Klein Centre for Cosmoparticle Physics,
	Department of Physics,
	Stockholm University,
	AlbaNova University Center,
	Roslagstullsbacken 21,
	SE--106\.91 Stockholm,
	Sweden}

\date{\formatdate{\day}{\month}{\year}, \currenttime}

%%%%%%%%%%%%%%%%%%%%%%%%%%%%%%%%%%%%%%%%%%%%%%%%%%%%%%%%%%%%

\begin{abstract}
\vs{-0.5mm}
Primordial black holes which are produced during an epoch of matter domination are expected to spin rapidly. It is shown that this leads to an enhancement of the detectability of the stochastic gravitational-wave background from their mergers. For a specific model, we explicitly demonstrate that this yields a $50\,\%$ increase of the gravitational-wave amplitude as compared to the non-spinning case.\\[-1.5mm]
\end{abstract}

\maketitle

%%%%%%%%%%%%%%%%%%%%%%%%%%%%%%%%%%%%%%%%%%%%%%%%%%%%%%%%%%%%
Gravitational-wave astronomy has become one of the most promising tools for modern observational cosmology and astrophysics. In particular, it may be able to ultimately discriminate the origin of observed merging black holes, \ie~to tell if these are of primordial origin \cite{ZeldovichNovikov69, Hawking:1971ei, Hawking:1974sw, Carr:1975qj}, and in this case, if they could constitute a significant part, or possibly all, of the dark matter (see Refs.~\cite{1975Natur.253..251C, GarciaBellido:1996qt, Carr:2009jm, Clesse:2015wea, Bird:2016dcv, Carr:2016drx, Clesse:2016vqa, Green:2016xgy, Garcia-Bellido:2017fdg, Kuhnel:2017pwq, Carr:2017jsz, Carr:2017edp, Clesse:2017bsw, Clesse:2018ogk, Carr:2018poi, Kuhnel:2018mlr, Carr:2019hud, Garcia-Bellido:2019vlf, Kuhnel:2019xes, Carr:2019kxo, Chen:2019zza}). In this situation, one expects a measurable stochastic gravitational-wave (GW) background from primordial black holes (PBHs) which have been merging over a large fraction of the cosmic history (cf.~Ref.~\cite{Clesse:2016vqa}). The characteristics of the associated GW signal depends on the specific mass distribution of the black holes, their velocities and spins.

Those properties are influenced by the environment within which the black holes form. Most of the literature discussing primordial-black hole formation focuses on gravitational collapse within a radiation-dominated area. However, several authors have entertained the possibility that the holes could have also been created during an epoch of matter-domination (see Ref.~\cite{Khlopov:1980mg, Polnarev:1982} for original references and Refs.~\cite{Harada:2016mhb, Carr:2018nkm, Carr:2017edp} for more recent discussions). It should be noted that such a phase may have exist before, and additionally to, the one after the matter-radiation equality at redshift $z \sim 10^{4}$, such as during an epoch after inflation with inflaton oscillations \cite{Assadullahi:2009jc, Alabidi:2009bk, Pi:2017gih, Martin:2019nuw}, or during the strong phase transition \cite{Khlopov:1980mg, Polnarev:1982, Sobrinho:2016fay}.

One crucial difference to a radiation environment is the absence of any pressure-gradient force in a matter background. Adopting the theory of angular momentum in structure formation \cite{Peebles:1969jm, Catelan:1996hv}, the authors of Ref.~\cite{Harada:2017fjm} recently found that rotation plays a very important r{\^o}le in the formation of primordial black holes within a matter-dominated area, and that most of these holes were already rapidly rotating at the time of their formation and will to a large extent continue to do so until now.

When those black holes merge, depending on how their spins are oriented to each other, the amount of energy emitted in gravitational waves will be either larger or smaller as compared to the non-spinning case, being maximal for aligned and minimal for anti-aligned spins \cite{Hemberger:2013hsa}.

In the {\it Letter}, we elaborate on the effect of PBH spin on the stochastic gravitational-wave background. This is demonstrated using a specific scenario in which the holes have been produced during a phase of matter domination. However, we have to stress that the obtained results are expected to hold in generic situations in which primordial black holes form during a sufficiently long period of matter domination, or have spun up significantly by other effects.

Below, we will utilise the scenario outlined in Ref.~\cite{Carr:2017edp}, consisting of two scalar fields which together determine the spectrum of perturbations: the inflaton field $\varphi$, which gives the dominant component to the curvature spectrum at large scales, and an additional light scalar ``spectator'' field $s$, which was energetically subdominant during the period of inflation, giving the dominant component at small scales on which it has a large amplitude.

Concretely, the total curvature power spectrum is assumed to be decomposed by
\begin{align}
	\Pcal_{\Rcal}
		&=
					\Pcal_{\Rcal,\.\varphi}
					+
					\Pcal_{\Rcal,\.s}
					\; ,
					\label{eq:PRSum}
\end{align} 
where the first (second) term is assumed to dominate on large (small) scales. The inflaton perturbations are taken to produce an almost scale-invariant spectrum in accordance with the observations of the cosmic microwave background, and shall assume the form
\begin{subequations}
\begin{align}
	\Pcal_{\Rcal,\.\varphi}( k )
		&=
					A_{\varphi}\mspace{-1mu}
					\left( 
						\frac{ k }{ k_{*} }
					\right)^{\!
						n_{\varphi}
						-
						1 
						+
						\frac{ 1 }{ 2 }
						\alpha_{\varphi}
						\ln
						\left( 
							\frac{ k }{ k_{*} }
						\right)
						}
					\; ,
					\label{eq:PRphi}
\end{align}
with the Planck pivot scale $k_{*} = 0.05\,{\rm Mpc}^{-1}$, the spectral index $n_{\varphi} = 0.968 \pm 0.006$, amplitude $A_{\varphi} \approx 2.14 \times 10^{-9}$, and running $\alpha_{\varphi} = - 0.0033 \pm 0.0074$ \cite{Akrami:2018odb}. The perturbation spectrum of the spectator field is assumed to have a similar form:
\begin{align}
	\Pcal_{\Rcal,\.s}( k )
		&=
					A_{s}\mspace{-1mu}
					\left( 
						\frac{k}{k_{*}}
					\right)^{\!
						n_{s}
						-
						1
						+
						\frac{ 1 }{ 2 }
						\alpha_{s}
						\ln
						\left( 
							\frac{ k }{ k_{*} }
						\right)
						}
					\; ,
					\label{eq:PRs}
\end{align}
\end{subequations}
with $A_{s} \ll A_{\varphi}$, $n_{s} > 1$ and $\alpha_{s} \leq 0$.

For the spectator-field model under consideration, the authors of Ref.~\cite{Carr:2017edp} derive an expression for primordial black-hole dark-matter fraction within the mass interval $( m_{\rm PBH},\.m_{\rm PBH} + \d m_{\rm PBH} )$ at the present epoch,
\vs{-1mm}
\begin{align}
	\psi( m_{\rm PBH} )\;\d m_{\rm PBH}
		&=
					A
					\left( 
						\frac{ m_{\rm PBH} }{ \mu }
					\right)^{\!-5\mspace{1mu}( n_{s} - 1 ) / 6}
					\label{eq:psidm}
					\\[1mm]
		&\phantom{=\;}
					\times
					\exp\!
					\left[
						\log^{2}\mspace{-4mu}
						\left( 
							\frac{ m_{\rm PBH} }{ \mu }
						\right)
					\right]
					\.
					\d m_{\rm PBH}
					\; ,
					\notag
\end{align}
which holds between the masses $M_{\rm min}$ and $M_{\rm max}$. These correspond to the smallest and largest scales which become non-linear during the matter-dominated era, respectively; outside of this interval, we have $\psi \equiv 0$. The amplitude $A$ can be fixed by requiring that the fraction of dark matter consisting of primordial black holes, $f_{\rm PBH} \equiv \rho_{\rm PBH} / \rho_{\rm DM}$, satisfies
\vs{-0.5mm}
\begin{align}
	f_{\rm PBH}
		&=
					\int_{M_{\rm min}}^{M_{\rm max}}\!\d m_{\rm PBH}\;\psi( m_{\rm PBH} )
					\; .
					\label{eq:fPBH}
\end{align}
The peak mass $\mu$ is given by \cite{Carr:2017edp}
\begin{align}
	\mu
		=
					\left( 
						\frac{ k_{\rm reh} }{ k_{*} }
					\right)^{\!3}
					\frac{
						M_{\rm Pl}^{3}\,T_{\rm reh}^{-2}
						}
						{ 
						\sqrt{
							\.\frac{ 16\pi^{3} }{ 45 }\.g_{*}( T_{\rm reh} )
						\,}
					}
					\; ,
					\label{eq:Mc}
					\\[-6.5mm]
					\notag
\end{align}
with $M_{\rm Pl} = \sqrt{1 / G\,}$ being the Planck mass, $G$ is Newton's constant, and $g_{*}( T_{\rm reh} )$ denotes the number of relativistic degrees of freedom, being a function of the reheating temperature $T_{\rm reh}$ (which should be higher than $4\,{\rm MeV}$ for not to interfere with big-bang nucleosynthesis \cite{Kawasaki:2000en, Hannestad:2004px, Ichikawa:2005vw, DeBernardis:2008zz}).

If the primordial black holes produced in this scenario constitute a significant fraction of the dark matter, frequent mergers will necessarily occur, emitting a spectrum of gravitational waves. The superposition of this radiation leads to an energy density $\rho_{\rm GW}$ and characteristic strain amplitude \cite{Phinney:2001di, Sesana:2008mz}
\vs{-1mm}
\begin{align}
	h_{\crm}( f )
		&=
					\sqrt{
						\frac{ 4\,G }{ \pi f^{2} }\.
						\frac{\d \rho_{\rm GW}}{\d\!\ln f}
					\,}
					\; ,
					\label{eq:hc}
					\\[-6.5mm]
					\notag
\end{align}
with $f$ being the observed gravitational-wave frequency.

The individual primordial black-hole capture rate has been derived in Ref.~\cite{Mouri:2002mc} using the criterion that for two passing black holes (in the following referred to as 'A' and 'B') in order to form a bound system, the energy loss through gravitational waves should be of the order of the kinetic energy. In the Newtonian approximation (which should be sufficiently accurate as the cross-section is much larger than the black-hole surface area), this leads to the capture rate \cite{Mouri:2002mc}
\vs{-1mm}
\begin{align}
\begin{split}
	\tau_{\rm PBH}^{\rm capt}( m_{\Arm},\.m_{\Brm} )
		&\approx
					4\.
					G^{2}\.
					c^{-10 / 7}\.
					n^{}_{\rm PBH}\,
					v_{\rm PBH}^{-11 / 7}
					\\[1mm]
		&\phantom{\approx\;}
		\times
					\!\big( m_{\Arm} + m_{\Brm} \big)^{10 / 7}\,
					\big( m_{\Arm}\.m_{\Brm} \big)^{2 / 7}\.
					\; ,
					\label{eq:capture_rate}
\end{split}
\end{align} 
with $n_{\rm PBH}$ being the primordial black-hole number density, and $v_{\rm PBH}$ is the relative velocity of the two holes.

Once bound, the merging of the two black holes typically happens on a time-scale of less than a million years. In turn, the authors of Ref.~\cite{Clesse:2016vqa} derive an expression for the merger rate per unit volume and per logarithmic interval of masses,
\begin{align}
	&\tau_{\rm merg}( m_{\Arm},\.m_{\Brm} )
		=
					\frac{
						\tau_{\rm PBH}^{\rm capt}( m_{\Arm},\.m_{\Brm} )\,
						\rho( m_{\Brm} )\.
						\rho_{\rm DM}\.
						f_{\rm DM}
						}
						{
						n_{\rm PBH}\,
						m_{\Arm}\.m_{\Brm}
					}
					\; ,
					\label{eq:taumerge}
\end{align}
where $\rho_{\rm DM}$ is the dark-matter density of the Universe. Applied to the spectator-field induced mass distribution Eq.~\eqref{eq:psidm}, we note that the merger rate Eq.~\eqref{eq:taumerge} has the qualitative behaviour
\begin{align}
\begin{split}
	\tau_{\rm merg}
		&\propto
					\exp\mspace{-3mu}
					\left[
						-
						\frac{ 5 }{ 36 }\.
						\alpha_{s}
						\left\{
							\log^{2}\!\left( \frac{ m_{\Arm} }{ \mu } \right)
							-
							\log^{2}\!\left( \frac{ m_{\Brm} }{ \mu } \right)\mspace{-2mu}
						\right\}
					\right]
					\\[2.5mm]
		&\phantom{\propto\;}
					\times
					\!\left(
						\frac{ m_{\Arm}\.m_{\Brm} }{ \mu^{2} }
					\right)^{\!2 / 7 - 5\mspace{1mu}( n_{s} - 1 ) / 6}\!
					\left(
						\frac{ m_{\Arm} + m_{\Brm} }{ \mu }
					\right)^{\!10 / 7}
					.
					\label{eq:taumerge2}
\end{split}
\end{align}

Following Ref.~\cite{Clesse:2016vqa}, integrating $\tau_{\rm merg}$ over the whole mass range and multiplying it by one half, one obtains the total merger rate, which can be used to derived the gravitational-wave amplitude
\begin{align}
	&\Omega_{\rm GW}( f )
		\propto
					f^{2/3}\.\int_{0}^{z_{\rm max}}
					\!\frac{\d z }{ H( z )\mspace{1mu}( 1 + z )^{4 / 3}}
					\label{eq:OmegaGW}
					\\[1.5mm]
	&\times
					\int \mspace{-10mu}\int\!
					\frac{\d m_{\Arm}}{m_{\Arm}}
					\frac{\d m_{\Brm}}{m_{\Brm}}\;
					\tau_{\rm merg}( m_{\Arm},\.m_{\Brm} )\!
					\left[
						\frac{ m_{\Arm}\.m_{\Brm}}
						{\,( m_{\Arm} + m_{\Brm} )^{1 / 3} }
					\right]^{5/3}
					,
					\notag
\end{align}
with the Hubble rate $H( z )$,
\begin{align}
\begin{split}
	H( z )^{2}
		&=
					H_{0}^{2}\,
					\big[
						( \Omega_{\rm DM} + \Omega_{\brm} )( 1 + z )^{3}
					\\[1mm]
		&\phantom{=\;H_{0}^{2}\,
					\Big[\mspace{-4.2mu}}
						+
						\Omega_{\rrm}( 1 + z )^{4}
						+
						\Omega_{\Lambda}
					\big]
					\, ,
					\label{eq:H}
\end{split}
\end{align}
where $H_{0}$ denotes the value of the Hubble rate today, and $\Omega_{\rm DM}$, $\Omega_{\brm}$, $\Omega_{\rrm}$, and $\Omega_{\Lambda}$ are the energy densities of the dark matter, the baryons, radiation and the cosmological constant, respectively, relative to the critical density of the Universe $\rho_{\crm}$.

The numerical prefactor in Eq.~\eqref{eq:OmegaGW} depends on the set of parameters $\{ f_{\rm PBH}, A_{s}, a_{\rm reh} / a_{\rm md}, n_{s}, T_{\rm reh} \}$, where $a_{\rm reh}$ and $a_{\rm md}$ are the scale factors at reheating and at the end of the matter-dominated phase, respectively. Exemplary, following Ref.~\cite{Carr:2017edp}, we will make the choice
\begin{equation}
\begin{alignedat}{3}
	f_{\rm PBH}
		&=
					1\; ,
						&\;\;
							n_{s}
								&=
									3.22\; ,
										&\;\;
											\alpha_{s}
												&=
													-\.
													0.131
					\; ,
					\\[1mm]
	\frac{ A_{s} }{ A_{\varphi} }
		&=
					10\; ,
						&\;\;
							T_{\rm reh}
								&=
									6\,{\rm MeV}\; ,
										&\;\;
											\frac{ a_{\rm reh} }{ a_{\rm md} }
												&=
													40
					\; .
\end{alignedat}
\end{equation}
This implies for the cut-off masses: $M_{\rm min} \approx 16.15\,M_{\odot}$ and $M_{\rm max} \approx 70.2\,M_{\odot}$. With $h \equiv H_{0} / 100\,{\rm km}\,{\rm s}^{-1}\,{\rm Mpc}^{-1}$, one finds
\begin{align}
	h^{2}\.\Omega_{\rm GW}( f )
		&\equiv
					8.4 \times 10^{-11}
					\left(
						\frac{ f }{ 1\,{\rm Hz} }
					\right)^{\!2/3}
					\; .
					\label{eq:hsquaredOmegaGW}
\end{align}

For the parameter choice mentioned above, the black holes produced by the spectator-field model under consideration are expected to have close to maximal spin \cite{Harada:2016mhb}. The authors of Ref.~\cite{Hemberger:2013hsa} have conducted a numerical study of black-hole mergers and found, for the case of both initial spins $\vec{\chi}_{i}$ being equal to and (anti-)aligned with each other, that the fraction of the black-hole energy radiated away though gravitational waves, $E_{\rm GW} \equiv 1 - M_{f} / M_{i}$, can be approximated by
\begin{align}
	E_{\rm GW}( \chi_{i} )
		&\approx
					0.00258
					-
					\frac{
						0.07730
						}
						{
						1.6939
						-
						\chi_{i}
					}
					\; ,
					\label{eq:Eem}
\end{align}
where $\chi_{i} \equiv \norm{\vec{\chi}_{i}}$, and $M_{f}$ is the mass of the final black hole after the merger and $M_{i}$ is the sum of the initial (Christodoulou) masses (see Ref.~[38] for details). So, despite the fact that the black holes are assumed to have maximal spin, depending on their relative orientation, the radiated energy varies. 

For arbitrarily oriented spins, we will conservatively assume that this behaviour holds for the components of the initial spin vectors $\vec{\chi}^{}_{i_{\Arm,\mspace{1mu}\Brm}}$ projected onto each other, while the orthogonal components are taken not to change the amount of radiated energy. Hence, averaging over the spin orientations, assuming an ensemble of randomly-oriented spinning black holes will lead to a surplus of radiation as compared to the non-spinning case. This is due to the non-linear nature of the emission in Eq.~\eqref{eq:Eem}. As this equation is only valid for both spins (anti-)aligned, an estimate for the gain as compared to the non-spinning case is $1 / 2\;[ E_{\rm GW}( 1 ) + E_{\rm GW}( - 1 ) ] / E_{\rm GW}( 0 )$.

\begin{figure}
	\vs{-4mm}
	\includegraphics[scale=1, angle=0]{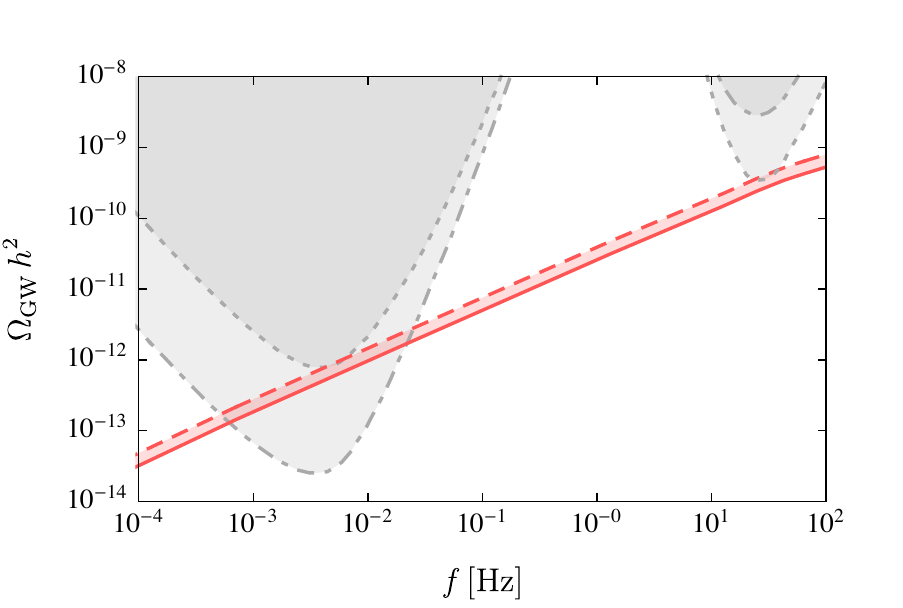}
	\caption{
			The stochastic gravitational-wave spectrum Eq.~\eqref{eq:hsquaredOmegaGW}
			for the spectator-field model of Ref.~\cite{Carr:2017edp},
			using exemplary $\mu = 12\,\Msun$, $v_{\rm vir} = 20\,{\rm km / s}$,
			and $f_{\rm PBH} = 1$.
			The red dashed curve shows the result including spin,
			while the red solid curves depicts the (hypothetical) case 
			in which the effect of spin has been neglected.
			The two dotted and dot-dashed grey curves at on the low-frequency side
			represent the expected sensitivities for LISA
			for the worst and best experimental designs, respectively \cite{Bartolo:2016ami}. 
			Also shown, on the high-frequency side, 
			are the limits from the O2 (dotted) and O5 run (dot-dashed, forecast)
			of Advanced LIGO {\cf~Fig.~3 of Ref.~\cite{Clesse:2016vqa}}).
			}
	\label{fig:Constraints}
\end{figure}

Applying this to the concrete spectator-field model yields an increase of gravitational-wave radiation of $50\.\%$. Figure \ref{fig:Constraints} displays the stochastic background amplitude $\Omega_{\rm GW}( f )$, which experiences an according amplification. This provides an explicit example of a scenario in which a large number of highly spinning black holes are produced, and it has been demonstrated how the stochastic gravitational-wave background signal from their merging is significantly enhanced.

While the effect of black-hole rotation discussed in this work is one amongst a number of others, such as clustering, eccentricity, multi-mergers, tidal forces, or accretion, which could possibly induce a large modification of a detectable gravitational-wave signal as well, it is necessary to know the size of the spin effect in order to pursue further the young but long road of precision gravitational-wave astronomy.
\vs{2mm}

%%%%%%%%%%%%%%%%%%%%%%%%%%%%%%%%%%%%%%%%%%%%%%%%%%%%%%%%%%%%

\acknowledgments
The author thanks Bernard Carr, Lawrence Kidder, and Bo Sundborg for stimulating discussion, and thanks the anonymous referee for valuable remarks. He acknowledges support from the Vetenskapsr{\aa}det (Swedish Research Council) through contract No.~638-2013-8993, and the Oskar Klein Centre for Cosmoparticle Physics.

%%%%%%%%%%%%%%%%%%%%%%%%%%%%%%%%%%%%%%%%%%%%%%%%%%%%%%%%%%%%

\bibliography{refs}

\end{document}